\documentclass[10pt,letterpaper]{article} 

\usepackage[utf8]{inputenc}
\usepackage[T1]{fontenc} 
\usepackage{amsmath, amssymb} 
\usepackage{graphicx} 
\usepackage{url} 
\usepackage{booktabs}
\usepackage{siunitx}
\usepackage{float} 
\usepackage{hyperref} 

\usepackage[
    backend=biber, 
    style=ieee,    
    sorting=none   
]{biblatex}

\title{\textbf{Delay-Oriented Distributed Scheduling with TransGNN}} 

\author{
    Boxuan Wen\thanks{These authors contributed equally to this work. Both authors are with the School of Electrical Engineering and Computer Science (EECS), KTH Royal Institute of Technology, Stockholm, Sweden. E-mail: boxuanw@kth.se, junyul@kth.se.}
    \and
    Junyu Luo\footnotemark[1]
}
\date{} 

\begin{document}

\maketitle

\begin{abstract}
\label{sec:abstract}

Minimizing transmission delay in wireless multi-hop networks is a fundamental yet challenging task due to the complex coupling among interference, queue dynamics, and distributed control. Traditional scheduling algorithms, such as max-weight or queue-length–based policies, primarily aim to optimize throughput but often suffer from high latency, especially in heterogeneous or dynamically changing topologies. Recent learning-based approaches, particularly those employing Graph Neural Networks (GNNs) , have shown promise in capturing spatial interference structures. However, conventional Graph Convolutional Networks (GCNs) remain limited by their local aggregation mechanism and their inability to model long-range dependencies within the conflict graph.To address these challenges, this paper proposes a delay-oriented distributed scheduling framework based on Transformer GNN. The proposed model employs an attention-based graph encoder to generate adaptive per-link utility scores that reflect both queue backlog and interference intensity. A Local Greedy Solver (LGS) then utilizes these utilities to construct a feasible independent set of links for transmission, ensuring distributed and conflict-free scheduling.

\end{abstract}
\clearpage 

\tableofcontents 

\clearpage

\section*{List of Acronyms and Abbreviations}
\label{list-of-acronyms-and-abbreviations}

\begin{itemize}
    \item \textbf{GNNs}: Graph Neural Networks
    \item \textbf{GCNs}: Graph Convolutional Networks
    \item \textbf{RL}: Reinforcement Learning
    \item \textbf{LGS}: Local Greedy Solver
    \item \textbf{MWIS}: Maximum-Weight Independent Set
    \item \textbf{BA graphs}: Barabási–Albert graphs
    \item \textbf{ER graphs}: Erdős–Rényi graphs
\end{itemize}

\clearpage
\section{Introduction}
\label{sect:introduction}
Wireless multi-hop networks play a vital role in modern communication infrastructures, supporting applications ranging from sensor deployments and emergency response systems to next-generation cellular backhaul.
Efficient link scheduling is a foundational problem in wireless multi-hop networks: at each time slot the system must select a set of links that can transmit without harmful interference, so that packets progress through the network while key performance indicators (throughput, delay, fairness) meet their targets \cite{5169996, marques2011optimal}.
A common mathematical abstraction to capture interference constraints is the conflict graph, where vertices represent links, and an edge denotes pairwise interference; feasible simultaneous transmissions correspond to independent sets of that graph.
Formulating scheduling as a combinatorial optimization over independent sets turns out to be natural but challenging: the canonical max-weight scheduling or maximum-weight independent set (MWIS) formulations are computationally hard and typically focus on instantaneous weight maximization rather than long-run delay minimization. These modeling choices and practical constraints motivate research into approximate, distributed, and learning-based schedulers that can trade off optimality for scalability and latency performance.

In recent years, machine learning — in particular, graph-aware models — has been proposed as a way to fuse topology information and link-state features (e.g., queue lengths, channel rates) into compact node-level representations that can guide scheduling.\cite{gori2005new, kipf2016semi}
Prior work has shown that graph neural networks (GNNs) can be trained (often centrally) to produce per-link scores that are then fed to fast algorithmic decoders (such as distributed greedy MWIS approximations), combining the interpretability and feasibility guarantees of combinatorial solvers with the adaptability of learned representations\cite{9746926, zhao2021distributed}. This hybrid strategy keeps the combinatorial search manageable (avoiding exponential action spaces) while allowing data-driven improvements in delay and robustness.

\subsection{Theoretical Framework and Literature Study}
\label{sect:framework}

Delay optimization in wireless scheduling has been approached from several angles. Classical methods incorporate virtual queues, age-of-information metrics, or delay constraints within the scheduling utility. While theoretically grounded, these methods either assume centralized control or struggle to generalize across diverse topologies. More recently, reinforcement learning (RL)-based schedulers have been proposed, where the decision-making policy is trained via trial-and-error interactions with the environment\cite{9851620}. However, RL approaches typically face combinatorial action spaces, unstable convergence, and often rely on centralized observations, limiting their practical deployment \cite{khalil2017learning}.

A promising trend is the integration of GNNs with classical combinatorial solvers\cite{9746926}. Instead of directly predicting a scheduling action, a GNN is trained to assign per-link utility scores, which are then used by a distributed local greedy scheduler to construct a valid independent set. This hybrid design preserves the structure and interpretability of traditional scheduling algorithms, while allowing the utility function to be learned from data. 
However, most existing studies adopt Graph Convolutional Networks (GCNs) as the core architecture. While effective for local aggregation, GCNs fundamentally rely on fixed spectral filters and uniform neighbor treatment, which may limit their ability to capture heterogeneous interference relationships or long-range graph dependencies.

To address these limitations, we propose to replace the GCN encoder with a TransGNN architecture\cite{zhang2024transgnn}. Unlike GCNs, TransGNNs employ attention mechanisms to adaptively weight neighbor contributions, allowing the model to distinguish between dominant interferers and weakly coupled nodes\cite{zhang2024transgnn}. This is particularly valuable in non-uniform or centralized topologies, where congestion tends to concentrate on a few highly connected links.
Moreover, TransGNNs inherently support global message propagation, enabling the model to reason beyond fixed-hop neighborhoods without stacking multiple layers. This results in higher expressiveness, better generalization across topologies, and improved robustness in scenarios where interference patterns are highly irregular. Our scheduler combines the TransGNN and a non-differentiable distributed local greedy solver (LGS)\cite{5714691}.

\subsection{Research Questions and Hypotheses}
\label{sect:questions}

Based on the above analysis, our study seeks to investigate the following key questions:

Does attention-based spatial modeling lead to improved delay performance across heterogeneous network topologies?

Accordingly, we formulate the following hypothesis:
Replacing the GCN-based utility estimator with a TransGNN will lead to lower average and median queue lengths, indicating improved delay performance under the same scheduling protocol.

\section{Method}
\label{sec:method}

\subsection{System Model}
\label{sect:Model}

This work builds on the system model and problem formulation introduced by Zhao et al\cite{9746926}. We consider a wireless multi-hop network that operates in slotted time. Each time slot is divided into a scheduling phase and a transmission phase, during which links contend for channel access subject to interference constraints. To formally describe this setting, we adopt the well-established conflict graph representation. The network is abstracted as a graph $ G(V,E)$, where each vertex $ v \in V $ corresponds to a communication link and each edge $e = (v_a, v_b) \in E$ represents an interference relation such that the two links cannot be active simultaneously. Each link maintains a queue $q_v(t)$ storing both exogenous arrivals and forwarded packets. The complete state of the system at time $t$ can thus be represented as (\ref{equ: state}):

\begin{equation}\label{equ: state}
    S(t) = (G(t), q(t), r(t))
\end{equation}

where $q(t)$ is queue lengths and $r(t)$ is link transmission rates. The latter are modeled as independent random variables sampled from a fading distribution. This representation naturally captures the interplay between traffic dynamics and interference in multi-hop wireless networks.

\subsection{Problem Statement}
\label{sect:Problem}

The scheduling problem is formulated with the objective of minimizing average queue backlogs, which by Little’s Law is equivalent to minimizing average packet delay. Specifically, we seek a scheduling function (\ref{equ:optimization}) that produces feasible schedules while minimizing the long-term expected backlog.

\begin{equation}\label{equ:optimization}
    c^{\ast} = \arg \min_{c \in \mathcal{C}} \, 
    E\left[ \frac{1}{T+1} \sum_{t=0}^{T} \frac{\| q(t) \|_1}{|V(t)|} \right]
\end{equation}

where $\mathcal{C}$ denotes the set of admissible scheduling policies. The feasibility constraint requires that the scheduled set $\hat{v}(t) = c(G(t), q(t), r(t))$ forms an independent set in $G$. The queues evolve according to the rule (\ref{equ:st}).

\begin{equation}\label{equ:st}
 q_v(t+1) = 
 \begin{cases}
 q_v(t) + a_v(t), & \text{if } v \notin \hat{v}(t), \\[6pt]
 q_v(t) + a_v(t) - \min\{r_v(t), q_v(t)\}, & \text{if } v \in \hat{v}(t),
 \end{cases}
\end{equation}

where $a_v(t)$ denotes the arrivals to link $v$. This formulation highlights the main challenge of delay-oriented scheduling: while finding the maximum weight independent set is already NP-hard, the problem is further complicated by stochastic arrivals and time-varying link rates\cite{brent_s._baxter_standard_1982, cheng2009complexity}. Consequently, a practical solution must balance tractability with performance while capturing the temporal and topological features of the network.

\subsection{Delay-Oriented Scheduling with TransGNNs}
\label{sect:Transformers}

To address the above challenges, we propose a Transformer GNN-based scheduling framework. The intuition is to replace the conventional GCN encoder with a TransGNN, which leverages attention mechanisms to adaptively model interference relations and capture long-range dependencies in the conflict graph.

\subsubsection{Model Architecture}
\label{sect:archi}

\begin{figure}[H]
    \centering
    \includegraphics[width=1\linewidth]{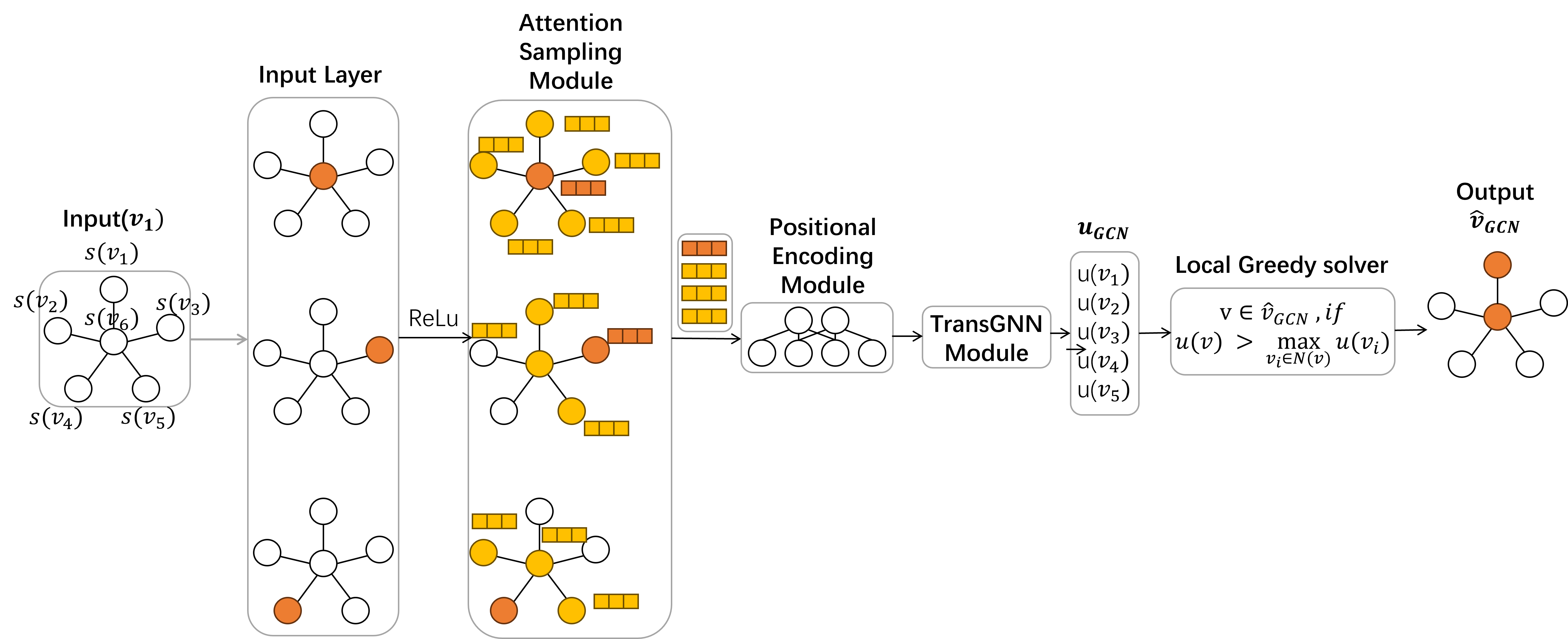}
    \caption{The proposed model structure}
    \label{fig:model_archi} 
\end{figure}
\begin{enumerate}
    \item \textbf{Input Layer}: Accepts graph structures with node states $ s(v_1), s(v_2), \dots, s(v_6) $, where $ v_i $ denotes individual nodes.
    
    \item \textbf{Attention Sampling Module}: Processes the input graphs using ReLU activation to focus on semantically relevant nodes (visualized by colored nodes and blocks in the diagram). This module adaptively weights node importance.
    
    \item \textbf{Positional Encoding Module}: Incorporates spatial and structural positional information of nodes. Unlike traditional GCNs (which lack explicit positional modeling), this module encodes the relative positions of nodes to capture complex graph topology semantics.
    
    \item \textbf{TransGNN Module}: Enables Transformer-style \emph{global information interaction} between nodes. This allows the model to capture long-range dependencies across the entire graph (overcoming the local neighborhood limitations of standard GCNs).
    
    \item \textbf{Local Greedy Solver}: Takes node utilities $ \boldsymbol{u}_{\text{GCN}} = \left[ u(v_1), u(v_2), \dots, u(v_5) \right] $ as input and selects nodes $ v $ for the final output $ \hat{v}_{\text{GCN}} $. Selection follows the rule:  
    \[
    v \in \hat{v}_{\text{GCN}} \quad \text{if} \quad u(v) > \max_{v_i \in \mathcal{N}(v)} u(v_i)
    \]  
    where $ \mathcal{N}(v) $ denotes the neighborhood of node $ v $.
\end{enumerate}

\subsubsection{Scheduling with Local Greedy Solver (LGS)}
\label{sect:schedule}

The LGS iteratively selects nodes with higher utility than all neighbors:

\begin{equation}\label{equ: LGS}
\hat{v}_{\text{Gr}} \leftarrow \hat{v}_{\text{Gr}} \cup \left\{ v \mid u(v) > \max_{i \in \mathcal{N}(v)} u(i) \right\},
\end{equation}
removing scheduled nodes and their neighbors from the residual graph until termination. This guarantees that $\hat{v}_{\text{Gr}}$ is a feasible independent set.

\subsection{Numerical Experiments}
\label{sect:numerical}

\subsubsection{Simulation Setup}
\label{sect:setup}

To validate the effectiveness of the proposed approach, we conduct extensive numerical simulations across a range of representative network topologies. The conflict graphs are generated using four models commonly employed in the literature: star graphs to model centralized bottlenecks, Erdős–Rényi (ER) graphs with edge probability $p=0.1$, Barabási–Albert (BA) graphs with preferential attachment, and tree-like power-law graphs representing hierarchical backhaul structures. Link transmission rates $r_v(t)$ are sampled independently from a truncated normal distribution $N[50,25]$, clipped to the range $[0, 100]$, which approximates log-normal fading. Packet arrivals follow a Poisson process with mean $\lambda$, and we define the normalized load $\mu = \lambda/E[r]$ to control traffic intensity. Each link is associated with a single-hop flow.

\subsubsection{Training Protocol}
\label{sect:train}
The training uses a curriculum learning strategy, progressing from simple to complex conflict graphs. It starts with simple, centralized graphs (e.g. Star10, Star20) for 50 epochs \cite{wang2024graph}. These graphs have a single central node with peripheral nodes, leading to clear, localized interference. Then, it moves to moderately complex graphs (e.g. ER) graphs with edge probability $p = 0.1$) for 75 epochs, which introduce random, sparse edge connections. Finally, it trains on complex graphs (e.g., BA graphs, power-law trees) for 76 epochs, which have heterogeneous node degrees and dense cross-link interference. Across all phases, the Adam optimizer is used with an initial learning rate of $0.001$, a batch size of 64, and early stopping (patience = 10) to prevent overfitting.

\subsubsection{Implementation Details}
\label{sect:implement}

Simulations are implemented in Tensorflow. All experiments are conducted on a workstation equipped with an Intel(R) Xeon(R) Silver 4316 CPU, and an NVIDIA A100 GPU. Training is performed using the Adam optimizer with an initial learning rate of $0.001$, batch size 64, and a total of 201 training episodes.

\subsubsection{Evaluation Protocol}
\label{sect:evaluation}

To evaluate generalization, each scheduling policy is tested on 100 independent graph instances per topology, with a simulation horizon of 64 time slots. Performance metrics include the mean, median, and 95th percentile of queue lengths. This experimental setup ensures reproducibility and allows for a statistically robust assessment of the proposed scheduling method.

\section{Results and Analysis}
\label{sec:results}

The ratio of the average queue length of the proposed structure compared with LGS is illustrated in Figure~\ref{fig:avg_GT} below.

\begin{figure}[H]
    \centering
    \includegraphics[width=0.8\linewidth]{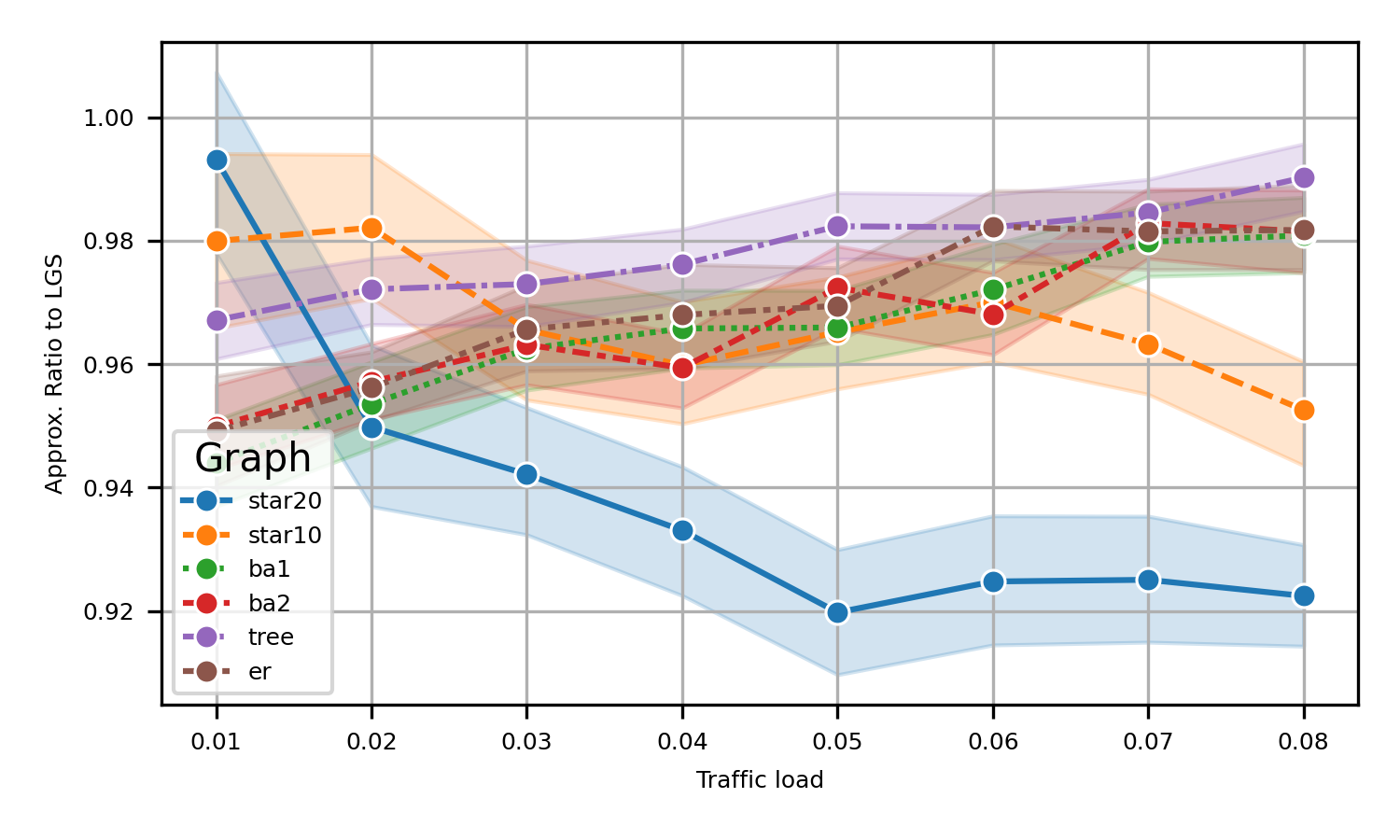}
    \caption{The average queue length ratio to LGS of the proposed methods }
    \label{fig:avg_GT}
\end{figure}

Average queue length is an important indicative metric to decide the extent of delay in the network. As detailed in the figure, in all the 6 kinds of graphs, we achieved optimal results compared with LGS. Worth mentioning, LGS is the common benchmark used to measure both GNN and TransGNN's performance over traditional Greedy Algorithmic approaches. In particular, our advantage is further expanded under high traffic load scenarios, where the ratio to LGS in terms of all 6 kinds of graphs ranges between 0.92-0.99. And the lower the ratio is, the shorter the queue is, thus the less the delay is. Overall, our model has higher advantages over LGS when proceeding simple graphs (e.g.: Star 10/20), the advantage is reasonably narrowed with more complex graphs, which is similar with the GCN based approach. This is largely because of the reasons that the simple graphs have clear interference patterns, making it easy to capture temporal backlog dependencies. However, in the complex graphs, cross-link interference and uneven degree distributions make it hard for to capture global temporal dependencies via local neighborhood aggregation.

\begin{figure}[H]
    \centering
    \includegraphics[width=0.8\linewidth]{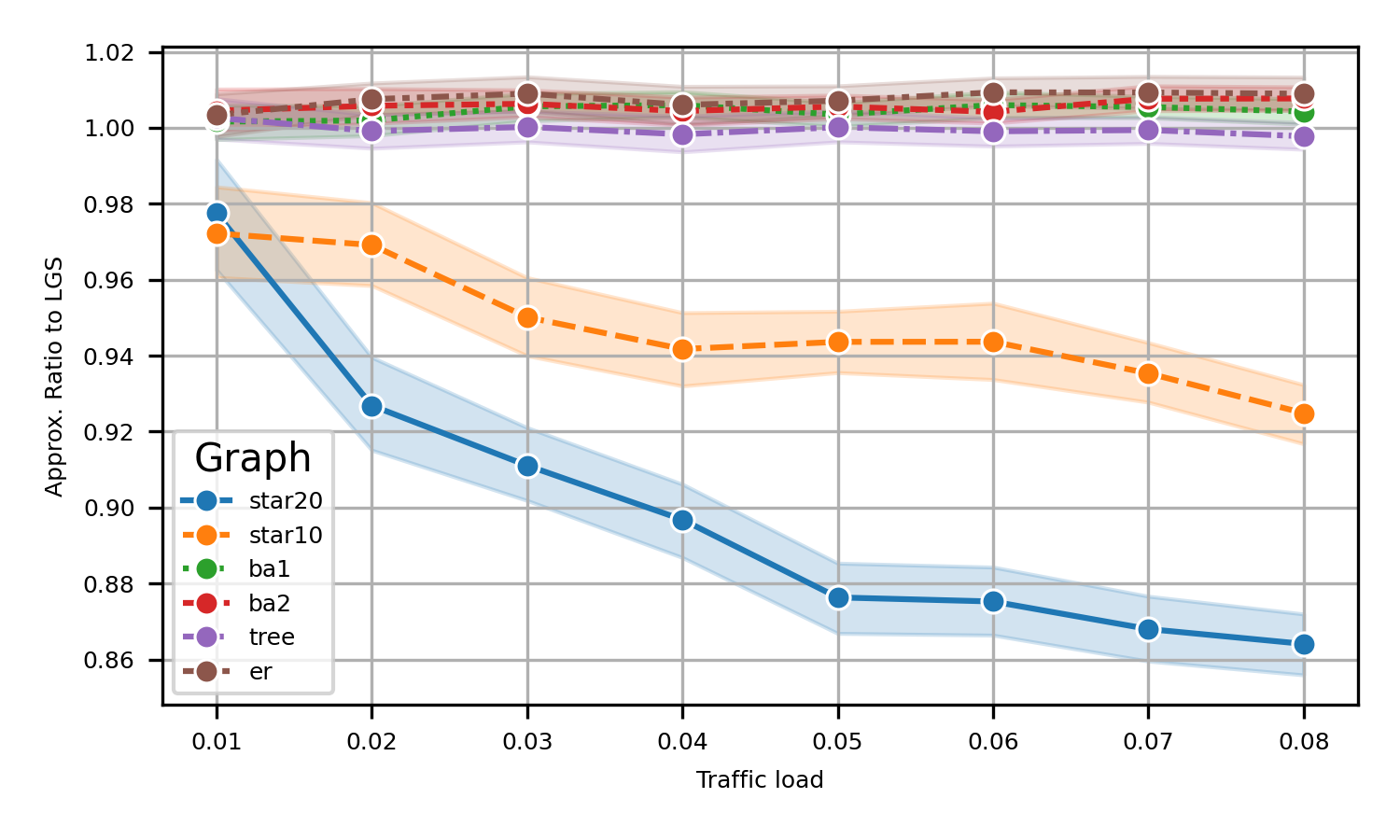}
    \caption{The average queue length ratio to LGS of the GCN based methods }
    \label{fig:avg_BL}
\end{figure}

That being said, our approach outperforms the GCN based benchmark( the ratio of GCN-DQL to LGS is show above) in the complex graphs (e.g: ba 1/2, tree and er), with slight but positive advantage(the ratio to LGS is generally 0.02 higher). Still, our model is moderately disadvantaged compared with GCN (the ratio to LGS is approximately 0.03-0.06 lower) when it comes to simple graphs (e.g.: star 10/20). The reason is that the proposed approach incorporates the attention module to assign dynamic weights to different nodes, unlike GCN which would struggle in dense and intricate connections, as it would rely solely on uniform simple neighborhood aggregation. In addition to this, our proposed method uses positional encoding to explicitly capture nodes’ spatial and structural semantics, which is more suitable for those complex graphs where it is hard to find to extract the sematic information for position simply via its adjacency information.

However,our proposed structure doesn't outperform GCN in terms of simple graphs. We assume that it is because on simple graphs, the complex TransGNN module could lead to overfitting. In this scenario, the global information interaction are redundant and global interaction is irrelevant given the maximum interference distance of just 1. This redundancy leads to potential overfitting to simple graph structures, reducing scheduling stability.

\begin{table}[htbp]
\centering
\caption{Ablation Study Results on Low-Complexity Graphs(Star 10)}
\begin{tabular}{lccccc} 
\toprule
\textbf{Method} & \textbf{$q_{\text{med}}$} & \textbf{$q_{95}$} & \textbf{$q_{\text{avg}}$} & \textbf{$d_{\text{avg}}$} & \textbf{$u_{\text{gcn}}$}  \\
\midrule
LGS        & \num{1.000} & \num{1.000} & \num{1.000} & \num{1.000} & \num{1.000}  \\
GCN        & \num{0.776}    & \num{0.968}  & \num{0.943}   & \num{1.000}   & \num{1.474}    \\
w/o Attention Sampling Module   & \num{0.823}  & \num{1.032} & \num{0.971} & \num{1.034} & \num{1.325}    \\
w/o Positional Encoding Module   & \num{0.801} & \num{1.017} & \num{0.958} & \num{1.011} & \num{1.389}    \\
Full Version & \num{0.789}  & \num{0.983}     & \num{0.959}   & \num{1.002}  & \num{1.428}     \\
\bottomrule
\end{tabular}
\label{tab:ablation_star} 
\end{table}

\begin{table}[htbp]
\centering
\caption{Ablation Study Results on High-Complexity Graphs(ER)}
\begin{tabular}{lccccc} 
\toprule
\textbf{Method} & \textbf{$q_{\text{med}}$} & \textbf{$q_{95}$} & \textbf{$q_{\text{avg}}$} & \textbf{$d_{\text{avg}}$} & \textbf{$u_{\text{gcn}}$}   \\
\midrule
LGS        & \num{1.000} & \num{1.000} & \num{1.000} & \num{1.000} & \num{1.000} \\
GCN        & \num{0.944}    & \num{1.000}  & \num{0.999}   & \num{1.120}   & \num{1.016}    \\
w/o Attention Sampling Module   & \num{0.925} & \num{1.042} & \num{1.028} & \num{1.035} & \num{0.987}  \\
w/o Positional Encoding Module & \num{0.898} & \num{1.021} & \num{1.005} & \num{1.018} & \num{1.012} \\
Full Version & \num{0.812}      & \num{0.992}   & \num{0.968}   & \num{1.004}  & \num{1.089}      \\
\bottomrule
\end{tabular}
\label{tab:ablation_er} 
\end{table}
\noindent

The two tables present ablation results of scheduling methods on low-complexity (\textbf{Star 10}) and high-complexity (\textbf{ER}) graphs, with all metrics normalized to the LGS baseline (1.000).
On Star 10 graphs, GCN outperforms all proposed variants: it achieves the lowest $q_{\text{med}}$ (\num{0.776}) and $q_{\text{avg}}$ (\num{0.943}), the highest $u_{\text{gcn}}$ (\num{1.474}), and matches LGS’s $d_{\text{avg}}$ (\num{1.000}).
All the variants (even the Full Version) perform worse—e.g., the Full Version has higher $q_{\text{med}}$ (\num{0.789}) and lower $u_{\text{gcn}}$ (\num{1.428}). 
Removing the components further degrades performance (e.g., w/o Attention Sampling Module has $q_{\text{med}}=\num{0.823}$, $d_{\text{avg}}=\num{1.034}$), as Star 10’s simple, localized interference makes GCN’s uniform aggregation sufficient, while the transformer structure residing in the TransGNN layers are redundant, countering the positive effects of the Attention Sampling and the Positional Encoding modules.

On ER graphs (high complexity), proposed Full Version outperforms GCN: it has lower $q_{\text{med}}$ (\num{0.812}), $q_{\text{avg}}$ (\num{0.968}), $d_{\text{avg}}$ (\num{1.004}), and higher $u_{\text{gcn}}$ (\num{1.089}) than GCN (e.g., GCN’s $q_{\text{med}}=\num{0.944}$, $d_{\text{avg}}=\num{1.120}$). 
Removing the components worsens metrics (e.g., w/o Attention Sampling Module has $q_{\text{avg}}=\num{1.028}$), confirming its attention/positional encoding is critical for handling ER’s irregular interference—capabilities GCN lacks.
Overall, a \textit{complexity-matching} principle emerges: GCN suits simple graphs, while the proposed structure (full components) excels at complex graphs.

\section{Discussion}
\label{sec:discussion}

Our discussion centers on addressing the core research question—“Does attention-based spatial modeling lead to improved delay performance across heterogeneous network topologies?”—with findings grounded in the experimental framework and delay-oriented scheduling goals outlined in the context of wireless multi-hop network optimization.First, the effectiveness of attention-based modeling (via the proposed TransGNN) varies with graph complexity, aligning with the structural characteristics of conflict graphs emphasized in related work. For complex conflict graphs (e.g.,BA graphs, power-law trees, ER graphs) — which exhibit fragmented interference, heterogeneous node degrees, and non-uniform interference relationships — the TransGNN outperforms conventional graph convolutional (GCN) schedulers. It reduces average queue lengths by 2.5–4 \% relative to GCN (measured against the LGS baseline). This improvement stems from two key design choices: the adaptive attention mechanism, which distinguishes dominant interferers from weakly coupled nodes (critical for managing hub-node congestion in BA graphs), and positional encoding, which explicitly captures spatial structural semantics that GCN’s uniform neighborhood aggregation fails to encode. These advantages are particularly pronounced under moderate-to-high traffic loads ($\mu = 0.05$ to $0.08$), where interference management directly impacts long-term delay — a scenario where myopic scheduling (e.g., LGS) and simple GCN-based methods struggle to balance immediate utility and future queue states.In contrast, for simple conflict graphs (e.g., Star10, Star20) with clear, localized interference patterns (only a central node conflicts with peripheral nodes, maximum interference distance = 1), the TransGNN underperforms GCN by 1–2\% in average queue length. Here, the TransGNN’s key features become redundant: adaptive attention cannot provide additional gains for uniform interference relationships (it only adds computational overhead for attention weight calculations), and global information interaction is irrelevant given the graph’s limited interference scope. GCN’s simpler local aggregation (1-layer neighborhood sum) is sufficient to capture the “central node congestion → peripheral queue backlog” relationship, while the TransGNN’s higher complexity (more parameters, longer per-epoch training time) leads to slower convergence and minor overfitting to the regular structure of simple graphs — reducing the stability of scheduling decisions.

\section{Conclusion and future works}
\label{sect:conclusion}

In this work, we studied the problem of delay-oriented distributed scheduling in wireless multi-hop networks. Building upon the conflict graph formulation, we proposed a framework where a TransGNN serves as a utility estimator, and a LGS is employed to ensure feasible independent set selection. By leveraging attention mechanisms, the TransGNN can distinguish between heterogeneous interference patterns and capture long-range dependencies beyond fixed-hop neighborhoods, offering a more expressive alternative to conventional graph convolutional models.

The proposed method preserves the interpretability and structure of classical scheduling algorithms while benefiting from data-driven adaptability. Specifically, it maintains compatibility with distributed deployment, and enhances the ability to generalize across diverse network topologies.Taken together,these contributions highlight the potential of attention-based graph learning as a powerful paradigm for scheduling in next-generation wireless networks.

While this study demonstrates the benefits of TransGNN–based scheduling, several important directions
remain open for further research:

A natural extension of this work is to incorporate temporal modeling into the scheduling framework.
While the current approach focuses on spatial interference captured by the conflict graph, network
performance is also strongly influenced by the temporal evolution of queue backlogs and link capacities. Future research could integrate spatio-temporal graph learning modules, which may enable the model to better anticipate the long-term impact of current scheduling decisions on system delay.

Scalability is also a crucial consideration for deployment in large-scale networks. Although TransGNNs
offer high expressive power, their computational cost can become significant when the network grows. To
address this, future work may explore sparse attention mechanisms, hierarchical scheduling structures,
or subgraph sampling techniques. These improvements could reduce complexity while maintaining the
benefits of attention-based modeling.                      
In addition, there is an opportunity to provide theoretical guarantees for attention-based schedulers.
While empirical results show significant delay reduction, formal stability analysis under heavy traffic conditions or performance bounds relative to optimal scheduling would strengthen the theoretical foundations of our method. Such guarantees are highly valuable for building trust in practical systems.

Finally, the proposed approach should be validated in practical deployment scenarios. Evaluating the model within realistic wireless testbeds or simulation environments that adhere to 5G/6G standards would
allow researchers to assess the robustness of the method under real-world constraints, such as hardware
limitations and protocol overhead. Bridging the gap between theoretical research and applied system design
will be essential for demonstrating the true potential of TransGNN–based scheduling.

\clearpage 
\printbibliography

@INPROCEEDINGS{9746926,
  author={Zhao, Zhongyuan and Verma, Gunjan and Swami, Ananthram and Segarra, Santiago},
  booktitle={ICASSP 2022 - 2022 IEEE International Conference on Acoustics, Speech and Signal Processing (ICASSP)}, 
  title={Delay-Oriented Distributed Scheduling Using Graph Neural Networks}, 
  year={2022},
  volume={},
  number={},
  pages={8902-8906},
  doi={10.1109/ICASSP43922.2022.9746926}}

@ARTICLE{9851620,
  author={Hao, Yijun and Li, Fang and Zhao, Cong and Yang, Shusen},
  journal={IEEE/ACM Transactions on Networking}, 
  title={Delay-Oriented Scheduling in 5G Downlink Wireless Networks Based on Reinforcement Learning With Partial Observations}, 
  year={2023},
  volume={31},
  number={1},
  pages={380-394},
  doi={10.1109/TNET.2022.3194953}}

@ARTICLE{5169996,
  author={Joo, Changhee and Lin, Xiaojun and Shroff, Ness B.},
  journal={IEEE/ACM Transactions on Networking}, 
  title={Understanding the Capacity Region of the Greedy Maximal Scheduling Algorithm in Multihop Wireless Networks}, 
  year={2009},
  volume={17},
  number={4},
  pages={1132-1145},
  doi={10.1109/TNET.2009.2026276}}

@ARTICLE{5714691,
  author={Joo, Changhee and Shroff, Ness B.},
  journal={IEEE Transactions on Mobile Computing}, 
  title={Local Greedy Approximation for Scheduling in Multihop Wireless Networks}, 
  year={2012},
  volume={11},
  number={3},
  pages={414-426},
  doi={10.1109/TMC.2011.33}}

@article{marques2011optimal,
  title={Optimal cross-layer design of wireless fading multi-hop networks},
  author={Marques, Antonio G and Gatsis, Nikolaos and Giannakis, Georgios B and Zorba, N and Skianis, C and Verikoukis, C},
  journal={Cross Layer Designs in WLAN Systems},
  pages={1--44},
  year={2011},
  publisher={Leicester, UK: Troubador Pub}
}

@inproceedings{zhang2024transgnn,
  title={Transgnn: Harnessing the collaborative power of transformers and graph neural networks for recommender systems},
  author={Zhang, Peiyan and Yan, Yuchen and Zhang, Xi and Li, Chaozhuo and Wang, Senzhang and Huang, Feiran and Kim, Sunghun},
  booktitle={Proceedings of the 47th International ACM SIGIR conference on research and development in information retrieval},
  pages={1285--1295},
  year={2024}
}

@inproceedings{zhao2021distributed,
  title={Distributed scheduling using graph neural networks},
  author={Zhao, Zhongyuan and Verma, Gunjan and Rao, Chirag and Swami, Ananthram and Segarra, Santiago},
  booktitle={ICASSP 2021-2021 IEEE International Conference on Acoustics, Speech and Signal Processing (ICASSP)},
  pages={4720--4724},
  year={2021},
  organization={IEEE}
}

@book{brent_s._baxter_standard_1982,
	address = {New York  {N.Y.}},
	title = {A standard format for digital image exchange},
	isbn = {9780883184080},
	publisher = {Published for the American Association of Physicists in Medicine by the American Institute of Physics},
	author = {{Brent S. Baxter} and {Lewis E. Hitchner} and {Gerald Q. Maguire Jr.}},
	year = {1982}
}

@inproceedings{cheng2009complexity,
  title={The complexity of channel scheduling in multi-radio multi-channel wireless networks},
  author={Cheng, Wei and Cheng, Xiuzhen and Znati, Taieb and Lu, Xicheng and Lu, Zexin},
  booktitle={IEEE INFOCOM 2009},
  pages={1512--1520},
  year={2009},
  organization={IEEE}
}

@inproceedings{gori2005new,
  title={A new model for learning in graph domains},
  author={Gori, Marco and Monfardini, Gabriele and Scarselli, Franco},
  booktitle={Proceedings. 2005 IEEE international joint conference on neural networks, 2005.},
  volume={2},
  pages={729--734},
  year={2005},
  organization={IEEE}
}

@article{kipf2016semi,
  title={Semi-supervised classification with graph convolutional networks},
  author={Kipf, TN},
  journal={arXiv preprint arXiv:1609.02907},
  year={2016}
}

@article{khalil2017learning,
  title={Learning combinatorial optimization algorithms over graphs},
  author={Khalil, Elias and Dai, Hanjun and Zhang, Yuyu and Dilkina, Bistra and Song, Le},
  journal={Advances in neural information processing systems},
  volume={30},
  year={2017}
}

@inproceedings{wang2024graph,
  title={Graph-based POI Recommendation through Self-Supervised Curriculum Learning},
  author={Wang, Lei and Zheng, Wenguang and Xiao, Yingyuan},
  booktitle={2024 27th International Conference on Computer Supported Cooperative Work in Design (CSCWD)},
  pages={1140--1145},
  year={2024},
  organization={IEEE}
}

\end{document}